\documentclass{ringb99}
\usepackage{graphics}
\begin{document}
\title{Current Status of Radio Observations of selected Clusters of Galaxies with the 100--m Telescope Effelsberg}
\author{Michael Thierbach\inst{1} \and Ulrich Klein\inst{2} \and Richard Wielebinski\inst{1}}  
\authorrunning{M. Thierbach~et al.}
\titlerunning{Current Status of Radio Observations of Galaxy Clusters with the Telescope Effelsberg}
\institute{Max--Planck--Institut f\"ur Radioastronomie,
 Auf dem H\"ugel 69, 53121 Bonn, Germany
\and Radioastronomisches Institut der Universit\"at Bonn, Auf dem H\"ugel 71, 53121 Bonn, Germany}
\maketitle

\begin{abstract}
We present the current status of radio observations of several clusters of galaxies done with the 100--m telescope Effelsberg. The purpose of this still ongoing project is to obtain new sensitive single-dish measurements of large extended diffuse radio sources in these clusters. The use of a single-dish is required to recover all the flux.

In this paper we present the current maps of the clusters Coma, Abell~2256 and Abell~2163.
\end{abstract}

\section{Introduction}
During the last months we used the 100--m telescope Effelsberg to map several clusters of galaxies known to harbor large extended diffuse radio sources. The still ongoing project is focused on the clusters Coma and Abell~2256.

Deiss~et~al.~(1997) presented a 1.4~GHz map of the diffuse radio halo of the Coma cluster done with the Effelsberg telescope with point sources subtracted. They also compare the integrated flux densities from Coma~C (the halo source) available in the literature. The high--frequency data points at 2.7 and 4.85~GHz (Schlickeiser~et~al.~1987; Waldthausen~1980) suggest a strong steepening of the spectrum. The aim of our measurements of Coma~C is to settle the open question of the steepening above 1.4~GHz, at least until 4.85~GHz.

In case of Abell~2256 we hope to verify the mini--halo visible in the map of Bridle~\&~Fomalont~(1976). Furthermore A2256, with its two giant relic sources, is an ideal laboratory for the study of the intracluster medium.

\section{Observations}
The observations were carried out at up to four frequencies (1.4, 2.7, 4.85, and 10.45~GHz). All receivers have cooled HEMT amplifiers. The 1.4~GHz system is located in the primary focus of the telescope, whereas the others are secondary focus systems. Their main properties can be found in Table~\ref{tabre}.

\begin{table}
      \caption{Receiver parameters.}
         \label{tabre}
         \begin{tabular}{lcrcc}
            \hline
            \noalign{\smallskip}
Receiver     & HPBW & Bandwidth & System      & Number of \\
$\nu\,$[GHz] &      &           & temperature & channels \\
            \noalign{\smallskip}
            \hline
            \noalign{\smallskip}
 1.4   & 9$\farcm 35$ &  20 MHz & 30~K & 2 \\
 2.675 & $\!\!\!$4$\farcm  3$ &  40 MHz & 40~K & 2 \\
 4.85  & $\!\!\!$2$\farcm  4$ & 500 MHz & 30~K & 4 \\
$\!\!\!$10.45  & 1$\farcm 15$ & 300 MHz & 50~K & 8 \\
            \noalign{\smallskip}
             \hline
         \end{tabular}
\end{table} 

At 1.4~GHz we mapped in right ascension and declination. At the other frequencies we used the azimuth as scan direction to be able to apply the CLEAN procedure (Klein~\&~Mack~1995) and in case of the 4.85 and 10.45~GHz systems to utilize the multibeam--technique. For the selected clusters, the details of the observations and the properties of the maps in the current stage of the project consult Table~\ref{tabcl}.

\begin{table*}
\caption{Observational parameters. ($^{\rm a}$ Integration time per independent point)}
         \label{tabcl}
         \hspace*{2.25cm}\begin{tabular}{lcrcccc}
            \hline
            \noalign{\smallskip}
Cluster & \multicolumn{2}{c}{Center of map} & Frequency & \multicolumn{1}{c}{Final} & \multicolumn{1}{c}{Integration} & rms-noise \\
& \multicolumn{1}{c}{$\alpha_{1950}$} & \multicolumn{1}{c}{$\beta_{1950}$} & & \multicolumn{1}{c}{map size} & \multicolumn{1}{c}{time$^{a}$} & \\
& [$\, ^h\quad ^m\quad ^s$] & [$\,\degr \quad\arcmin \quad\arcsec$] & [GHz] & \multicolumn{1}{c}{[$\,'\times \;'$]} & [sec] & [mJy/b.a.] \\
            \noalign{\smallskip}
            \hline
            \noalign{\smallskip}
Coma        & 12 54 40 &   27 55 00 & 2.675 & 150$\times$150 & 11 & 1.4 \\
Coma Center & 12 57 00 &   28 10 00 & 2.675 &  60$\times$60  & 28 & 1.0 \\
            & 12 57 12 &   28 14 00 & 4.850 &  57$\times$57  &  9 & 0.6 \\
A85         & 00 39 06 & $-$9 38 00 & 1.400 &  90$\times$90  &  5 & 5.0 \\
A262        & 01 49 54 &   35 55 00 & 1.400 & 120$\times$120 & 10 & 6.1 \\
A2163       & 16 12 54 & $-$6 00 00 & 1.400 &  72$\times$72  &  7 & 5.3 \\
            &          &            & 2.675 &  75$\times$75  &  9 & 1.5 \\
            &          &            & 4.850 &  43$\times$43  & 11 & 0.7 \\
A2218       & 16 35 42 &   66 20 00 & 1.400 &  72$\times$72  & 12 & 3.5 \\
            &          &            & 2.675 &  54$\times$54  & 10 & 1.6 \\
            &          &            & 4.850 &  48$\times$48  &  6 & 0.7 \\
A2255       & 17 12 30 &   64 07 00 & 1.400 &  72$\times$72  & 28 & 5.4 \\
            &          &            & 2.675 &  56$\times$56  &  8 & 2.7 \\
            &          &            & 4.850 &  28$\times$28  & 12 & 0.7 \\
A2256       & 17 06 00 &   78 45 00 & 1.400 &  72$\times$72  & 44 & 4.6 \\
            &          &            & 2.675 &  60$\times$60  & 11 & 1.9 \\
            &          &            & 4.850 &  36$\times$36  & 29 & 0.4 \\
            &          &    & $\!\!\!$10.450 & 30$\times$30 & 3.5 & 1.1 \\
A2319       & 19 19 30 &   43 52 00 & 1.400 &  72$\times$72  & 16 & 4.6 \\
            &          &            & 2.675 &  75$\times$75  &  5 & 3.2 \\
            &          &            & 4.850 &  47$\times$47  &  6 & 0.9 \\
            \noalign{\smallskip}
             \hline
         \end{tabular}
\end{table*} 

\section{The maps}

\begin{figure}[h]
\parbox[][9cm][]{\hsize}{}
\caption[]{Central region of the Coma cluster at 4.85~GHz. No significant polarization was detected.}
\label{coma6}
\end{figure}

\paragraph{Coma}
The 4.85~GHz map is seen in Fig.~\ref{coma6}. We observed the center of the cluster. Clearly one can see the central galaxies NGC~4869 and NGC~4874 surrounded by extended emission.

At 2.675~GHz we mapped a large area covering the central region of the Coma cluster and the extended source 1253+275 (see Fig.~\ref{coma11} for maps). For the center region we add a couple of additional coverages. The central extended emission (Coma~C) is obviously visible. The emission from the source 1253+275 shows strong polarization. Note that we do not detect the bridge connecting Coma~C to 1253+275 (Kim~et~al.~1989).

\begin{figure*}
\parbox[][11.5cm][]{0.6\hsize}{}

\vspace*{-12cm}

\hspace*{11cm}
\parbox[][7cm][]{0.36\hsize}{}

\vspace*{-0.5cm}

\hspace*{11cm}
\parbox[][6cm][]{0.36\hsize}{}
\caption[]{The Coma cluster of galaxies at 2.675~GHz. Left the full map (fig2a), top right the central region (fig2b), bottom right the source 1253+275 (fig2c). The polarized intensity is represented by the length of the E--vectors.}
\label{coma11}
\end{figure*}   

\begin{figure*}
\vspace*{-5mm}
\parbox[][7cm][]{0.33\hsize}{}
\parbox[][7cm][]{0.33\hsize}{}
\parbox[][7cm][]{0.33\hsize}{}
\caption[]{Abell~2163 maps: left 1.4~GHz (fig3a), middle 2.675~GHz (fig3b), right 4.85~GHz (fig3c).}
\label{2163}
\end{figure*}

\vspace{-2mm}\paragraph{Abell~85}
This cluster was observed at 1.4~GHz. The relic source southwest of the center is too small in extent to be resolved with the Effelsberg beam. Nevertheless an extension of emission towards the location of the relic is visible.

\vspace{-2mm}\paragraph{Abell~262}
This cluster was also mapped at 1.4~GHz. No extended emission is visible until now.

\vspace{-2mm}\paragraph{Abell~2163}
Herbig~\&~Birkinshaw~(1994) reported a discovery of a radio halo source of large size in this cluster. We observed this cluster at three frequencies. The maps (presented in Fig.~\ref{2163}) clearly show the existence of an extended source that does not have any significant polarized emission.

Feretti~(1999) also detected this halo source in her VLA observations.

\vspace{-2mm}\paragraph{Abell~2218}
For this cluster we have maps at three frequencies (1.4, 2.675, 4.85~GHz). Only at 4.85~GHz we can resolve the center. A large extended source is not visible. 

\vspace{-2mm}\paragraph{Abell~2255}
This cluster we also mapped at 1.4, 2.675 and 4.85~GHz. So far we can detect the halo candidate only in the 4.85~GHz map, after subtraction of the point-like sources.

\vspace{-2mm}\paragraph{Abell~2256}
All four frequencies were used to map this source (see Fig.~\ref{2256}). At present the 10.45~GHz map consists of combination of only 7 coverages (3.5 seconds of integration time per independent point), more coverages will be added in future. The large relic source north of the center is visible at 1.4~GHz as an extension of the point-like emission, becoming more distinct at the higher frequencies. The strong linear polarization of the extended source suggests that this is a relic structure, rather than a halo; halos usually do not exhibit strong linear polarization (Feretti~\&~Giovannini~1996).

Excess emission east of the center is visible in the three low frequency maps, which seems to be the mini--halo discussed by Bridle~\&~Fomalont~(1976).

\begin{figure*}
\vspace*{-2mm}
\hspace*{10mm}\parbox[][9cm][]{0.45\hsize}{}
\parbox[][9cm][]{0.45\hsize}{}
\end{figure*}
\begin{figure*}
\vspace*{-5mm}
\hspace*{10mm}\parbox[][9cm][]{0.45\hsize}{}
\parbox[][9cm][]{0.45\hsize}{}
\caption[]{Abell~2256: top left 1.4~GHz (fig4a), top right 2.675~GHz (fig4b), bottom left 4.85~GHz (fig4c), bottom right 10.45~GHz (fig4d). The vectors have the same meaning as in Fig.~\ref{coma11}. Note their Faraday rotation from low to high frequencies. This manifests the presence of a magneto--ionic intracluster medium. At 10.45~GHz the polarization data are still insufficient.}
\label{2256}
\end{figure*}

\vspace{-2mm}\paragraph{Abell~2319}
We mapped A2319 at the three low frequencies. Until now the halo source is only barely visible.

\section{Conclusion}
The observations, especially those of the Coma cluster and Abell~2256, present valuable new data. The extended sources are visible at various frequencies, so spectral ana\-lyses can be made in future.

In case of Abell~2163 we detect the halo source at 1.4, 2.675 and 4.85~GHz. The detailed study is in preparation.

\end{document}